\documentclass[%
onecolum,
 aip,
 jcp,%
 amsmath,amssymb,
reprint,%
]{revtex4-1}
\usepackage[ansinew]{inputenc}
\usepackage{graphicx}
\usepackage{dcolumn}
\usepackage{bm}
\usepackage{amssymb,amsmath,latexsym,amsfonts}
\usepackage{amsthm}
\usepackage{amsbsy}
\usepackage[mathscr]{eucal}
\usepackage[dvipdf]{epsfig}
\usepackage{color}
\usepackage[T1]{fontenc}

\newtheorem{proposition}{Proposition}

\newcommand{\CM}{\mathrm{CM}}
\newcommand{\MSD}{\mathrm{MSD}}

\newcommand{\eff}{\mathrm{eff}}
\newcommand{\diag}{\mathrm{diag}}

\begin{document}

\title{Non-reciprocal Rouse chain model}
\author{Leonardo Apaza}
\affiliation{Carrera de Ciencias F\'isicas, Universidad P\'ublica de El Alto, Bolivia}

\author{Mario Sandoval}
\email{sem@xanum.uam.mx}
\affiliation{Department of Physics, Complex Systems, Universidad Autonoma Metropolitana-Iztapalapa, Mexico City 09340, Mexico}

%

\begin{abstract}
We study the center-of-mass mean-squared displacement (MSD) of a free non-reciprocal Rouse chain. Three stochastic models are considered: an overdamped linear chain, an underdamped linear chain, and an underdamped nonlinear chain with finite natural bond length that fully takes into account the elastic interactions. For the overdamped model, an exact transient expression for the center-of-mass MSD is obtained from the associated Ornstein--Uhlenbeck process. We show that non-reciprocity  enhances the effective diffusion with respect to the classic Rouse model, and that it separates the translational right and left zero modes, so the arithmetic center of mass is not, in general, the dynamically decoupled translational coordinate. It is found that the normalized left zero mode gives a closed expression for the effective diffusivity in terms of the non-reciprocal parameter and the number of beads ($N$) which still scales as $N^{-1}$. Interestingly, the same long-time diffusivity coefficient is found for both, the underdamped linear model and for the less studied finite-natural-length nonlinear model. It is concluded that inertia and finite bond length, modify the transient dynamics of the system, while non-reciprocity determines the long-time diffusion coefficient.
\end{abstract}

\date{\today}
\maketitle

\affiliation{
 Department of Physics, Universidad Autonoma Metropolitana-Iztapalapa, 
 Mexico City 09340, Mexico.
}

\section{Introduction}
The Rouse model is the cornerstone of polymer dynamics~\cite{Rouse1953,DoiEdwards1986}. It has been useful at predicting relaxation times, diffusion, and viscoelastic properties of unentangled polymer melts. It is also the foundation of reptation theories \cite{Rubi1}. Since its introduction, the Rouse model has been extended by adding hydrodynamic interactions (Zimm model) \cite{DoiEdwards1986}, volume exclusion between particles, time-dependent crosslinks (Sticky Rouse model) \cite{Rubi2}, spatial and time dependence of stiffness and friction coefficients \cite{Sato}, nonlinear springs (FENE-P model) \cite{Peter}, topological constraints (tube model) \cite{Pier,DoiEdwards1986}, among others. More recently, the Rouse model  has been extended to active matter, where a generic active force was introduced and its effect on the chain's  diffusion was analytically found \cite{Osman}.  Force dipoles
have been also added, thus modeling the dynamics of the cell nucleus \cite{David}.  Other proposed active force dynamics is the addition of tangential forces to the Rouse chain. Its center of mass diffusion was also analytically calculated \cite{Peterson}. However and to our knowledge,  the  Rouse model has only been studied under the assumption of reciprocal interactions  (symmetry in the action-reaction forces) among beads.  Hence, the question we address in this work is: What would be the effect of non-reciprocal interactions in the the Rouse model diffusion? This non-reciprocal interaction asymmetry (also understood as a system's response dependence to the direction of which an input is applied) has been studied in several branches of physics (optics, \cite{Potton,optical}, electromagnetism \cite{Caloz}, quantum \cite{Peano,Ulla,Ulla2,quantum} and classical  mechanics \cite{Lisa,elastic}) for a long time. Quite recently,  non-reciprocal interactions have been discovered in soft matter systems such as in rotating magnetic colloids \cite{Shelley} and in spinning sea urchin embryos \cite{Tan}. Metamaterials \cite{elastic,Martin} research has also exploited this condition, see for example the fishbone metamaterial \cite{Coulais}, whose deformation is a function of the orientation of the applied external tension. Active metamaterials (in the sense of energy injection to the system) investigations, have also studied the effect of non-reciprocity \cite{Coulais,Coulais2,Sandoval}, and finally, works considering flying interacting spins as a model for flock behavior, have also seen the importance of adding non-reciprocal interactions \cite{Bandini,Cava4,Dadhichi,Chen}. 
 
 Motivated by the previous literature, in this paper we study a  free stochastic (subject to thermal fluctuations) bead--spring chain with an asymmetry in its action-reaction interaction forces, thus extending the Rouse model to the non-reciprocal case. We analytically and numerically focus on the diffusion of its arithmetic center of mass. To analyze this problem, we consider three related models: an underdamped nonlinear (due to the full interacting elastic force expression, and usually disregarded in the literature) chain with finite natural bond length, an underdamped linear chain obtained by setting $l_0=0$, and finally, an overdamped linear model. The latter is particularly useful because the bead coordinates form a multivariate Ornstein--Uhlenbeck process and the transient center-of-mass MSD can be obtained exactly from the one- and two-time second moments~\cite{gardiner2009stochastic,oksendal2003stochastic,da2014stochastic}. For the linear models, it is found that a bi-orthogonal mode decomposition gives the weighted translational coordinate and the corresponding effective diffusivity.  Interestingly, it is proven that the same long-time overdamped diffusivity is maintained in the underdamped models (even for the nonlinear case) as long as the difference between the arithmetic center of mass and the weighted translational coordinate has a displacement with uniformly bounded second moment. In the nonlinear case, this difference can be written directly in terms of the bond vectors, and the required bond confinement is proved for an introduced  regularized interaction force dynamics. The analytical results are finally compared to Brownian dynamics simulations. In particular, we test the exact transient MSD, the dependence of the effective diffusivity on non-reciprocity and chain length, and the common long-time slope for the three models.

%

The structure of the paper is as follows. Section~\ref{seccmodelo} introduces the three models and a numerical simulation of them provides their general behavior. Section~\ref{seccispectral} establishes the spectral properties of the linear drift matrix. Section~\ref{seccionmsd} derives the exact transient MSD from the one- and two-time second moments. Section~\ref{seccmode} presents the biorthogonal mode decomposition and the analytical long-time effective diffusion coefficient. Section~\ref{sectzeromodedom} states the asymptotic transfer principle used in the subsequent analysis. Section~\ref{secunderdampedloti} applies this result to the underdamped linear model and to the finite-$l_0$ nonlinear mode. Section~\ref{secnum} presents the numerical validation of the analytical results, and Sec.~\ref{secdiss} discusses the physical interpretation of the results. Technical derivations are collected in the Appendices.

\section{Model}
\label{seccmodelo}
\begin{figure}[h]
    \centering
    \includegraphics[width=0.6\linewidth]{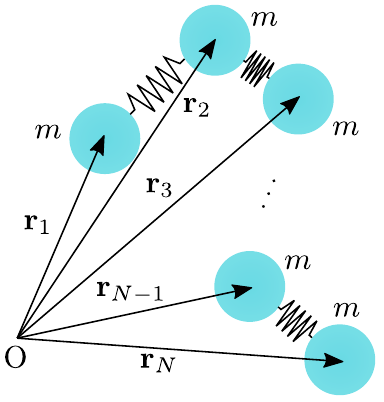}
    \caption{Free nonreciprocal bead--spring chain with nearest-neighbor interactions and free boundaries. The couplings transmitted across each bond are direction-dependent, with right and left stiffnesses $k_R$ and $k_L$, respectively.}
    \label{figmodel}
\end{figure}

Let us consider a chain of $N\ge2$ beads in $d$ spatial dimensions with nearest-neighbor interactions and free boundaries, as shown in Fig.~\ref{figmodel}. The position and velocity of bead $n$ are denoted by $\mathbf r_n(t)\in\mathbb R^d$ and $\mathbf v_n(t):=\dot{\mathbf r}_n(t)$, respectively. Here $m>0$ is the bead mass, $\gamma>0$ is the viscous friction coefficient, and $l_0\ge0$ is the natural bond length. We define the bond vectors as $\mathbf b_n(t):=\mathbf r_{n+1}(t)-\mathbf r_n(t)$, with $b_n(t):=\|\mathbf b_n(t)\|$, for $n=1,\dots,N-1$. Nonreciprocity is introduced through $k_L:=k(1+\delta)$ and $k_R:=k(1-\delta)$, with $k>0$ and $|\delta|<1$. Note that $\delta=0$ gives the reciprocal case $k_L=k_R$, while the additional limits $l_0=0$ and $m\to0$ recover the standard overdamped Rouse chain.

\subsection{Underdamped nonreciprocal chain with finite natural bond length}
Let us first consider the most general case, that is, an underdamped chain with finite bond length. Following the central-force law between two beads
\begin{equation}\label{eqcentralf1}
 \mathbf{f}_{S}: \mathbb{R}^d\setminus\{\mathbf{0}\} \to \mathbb{R}^d,
 \qquad
 \mathbf{f}_{S}(\mathbf{b}) := k_{S}\left(1-\frac{l_0}{\|\mathbf{b}\|}\right)\mathbf{b},
\end{equation}
where $S\in\{L,R\}$ labels the left or right coupling, and  on collision-free configurations, i.e., configurations satisfying $\mathbf b_j\neq\mathbf 0$ for all $j$, the internal force acting on bead $n$ is then given by
\begin{align}
 \mathbf{F}_1 &:= \mathbf{f}_R(\mathbf{b}_1),
 \label{eqmodel1f1}
 \\
 \mathbf{F}_n &:= \mathbf{f}_R(\mathbf{b}_n)-\mathbf{f}_L(\mathbf{b}_{n-1}),
 \qquad 2\le n\le N-1,
 \label{eqmodel1f2}
 \\
 \mathbf{F}_N &:= -\mathbf{f}_L(\mathbf{b}_{N-1}).
 \label{eqmodel1f3}
\end{align}
Hence the stochastic equations of motion for the whole chain become
\begin{align}
 \dot{\mathbf{r}}_n(t) &= \mathbf{v}_n(t),
 \\
 m\dot{\mathbf{v}}_n(t) &= \mathbf{F}_n(\{\mathbf{r}\}) - \gamma\mathbf{v}_n(t)
 + \sqrt{2\gamma k_{\text{B}} T}\,\boldsymbol{\eta}_n(t),
 \label{nsl}
\end{align}
where $\boldsymbol{\eta}_n(t)$ is Gaussian white noise with covariance $\langle \eta_{n\alpha}(t)\eta_{m\beta}(t')\rangle
 = \delta_{nm}\delta_{\alpha\beta}\delta(t-t')$. This model contains both inertia and nonlinear bond relaxation about the natural length $l_0$. Because the direction $\mathbf b_n/\|\mathbf b_n\|$ is undefined at $\mathbf b_n=\mathbf 0$, the above central force law  is understood to be defined on collision-free trajectories.  Nevertheless, a regularized short-distance force may be used in numerical work if bond crossings or bead  overlaps are allowed. To do so,  we use the smooth proportionality-preserving form
\begin{equation}\label{eqregularizedforce}
 \mathbf f_S^{(\varepsilon)}(\mathbf b_n)
 :=k_S\left(1-\frac{l_0}{\sqrt{\|\mathbf b_n\|^2+\varepsilon^2}}\right)\mathbf b_n,
 \qquad 0<\varepsilon\ll 1,
\end{equation}
for which $\mathbf f_R^{(\varepsilon)}=q\,\mathbf f_L^{(\varepsilon)}$ holds exactly. This regularized force is proven to be bounded
in Appendix~\ref{appconfinement}.

\subsection{Underdamped linear nonreciprocal chain}
\label{ssecunderdampedlin}
The linear underdamped model is obtained by setting $l_0=0$. The central-force law (Eq.~\eqref{eqcentralf1}) then reduces to $\mathbf{f}_{S}(\mathbf{b}) = k_{S}\mathbf{b}$, and the internal forces become linear in the coordinates (thus recovering the so-called underdamped Rouse model), namely
\begin{align}
 \mathbf{F}_1 &= k_R(\mathbf{r}_2-\mathbf{r}_1),
 \\
 \mathbf{F}_n &= k_R(\mathbf{r}_{n+1}-\mathbf{r}_n)-k_L(\mathbf{r}_n-\mathbf{r}_{n-1}),
 \,\,\, 2\le n\le N-1,
 \\
 \mathbf{F}_N &= -k_L(\mathbf{r}_N-\mathbf{r}_{N-1}).
\end{align}
For any Cartesian component, define the bead-coordinate vector $\mathbf{r}(t):=\left(r_1(t),\ldots,r_N(t)\right)^T$. Then the linear internal force can be written as $\mathbf{F}(\mathbf{r}):=-\mathbf{K}\mathbf{r}$, where the bead-index elastic matrix $\mathbf{K}$ is defined as
\begin{equation}
\mathbf{K}:=
\begin{pmatrix}
 k_R & -k_R & 0 & \cdots & 0 \\
 -k_L & k_L+k_R & -k_R & \ddots & \vdots \\
 0 & -k_L & k_L+k_R & \ddots & 0 \\
 \vdots & \ddots & \ddots & \ddots & -k_R \\
 0 & \cdots & 0 & -k_L & k_L
\end{pmatrix}.
\label{eqKmatrix}
\end{equation}
This matrix is the elastic operator of the linear nonreciprocal chain.

For this linear case, 
 each Cartesian coordinate is governed by the same non-symmetric operator $\mathbf K$, and hence the underdamped chain is an Ornstein--Uhlenbeck (OU) process in phase space $(\mathbf r,\mathbf v)$. Notice however that the configuration vector alone does not form a closed first-order OU process. 

\subsection{Non-inertial limit}
In the noninertial regime, the velocity relaxation time $m/\gamma$ is assumed to be much shorter than the configurational time scale, hence by taking the Smoluchowski--Kramers limit $m\to0$ to Eq.~\eqref{nsl} gives
\begin{equation}
 \gamma\dot{\mathbf{r}}_n(t)=\mathbf{F}_n(\{\mathbf{r}\})+\sqrt{2\gamma k_{\text{B}} T}\,\boldsymbol{\eta}_n(t).
 \label{equvecoverdlimit}
\end{equation}
For the linear force law, Eq.~\eqref{equvecoverdlimit} decouples componentwise because the deterministic force is linear and isotropic in Cartesian space and the Cartesian noise components are independent. It is therefore sufficient to let $r_n$ denote any Cartesian component of $\mathbf r_n$. The overdamped dynamics then reduces to the scalar equations
\begin{align}
 \gamma \dot{r}_1(t)
 &=
 k_R\bigl(r_2-r_1\bigr)
 +\sqrt{2\gamma k_{\text{B}} T}\,\eta_1(t),
  \label{over1}
 \\
 \gamma \dot{r}_n(t)
 &=
 k_R\bigl(r_{n+1}-r_n\bigr)
 -k_L\bigl(r_n-r_{n-1}\bigr)
 \nonumber\\
 &\quad
 +\sqrt{2\gamma k_{\text{B}} T}\,\eta_n(t),
 \qquad 2\le n\le N-1,
 \\
 \gamma \dot{r}_N(t)
 &=
 -k_L\bigl(r_N-r_{N-1}\bigr)
 +\sqrt{2\gamma k_{\text{B}} T}\,\eta_N(t),
 \label{over3}
\end{align}
where the scalar Gaussian white noises are normalized by $\langle \eta_n(t)\eta_m(t')\rangle = \delta_{nm}\delta(t-t')$.

\subsection{Overdamped linear nonreciprocal chain in matrix form}
For a generic Cartesian component, using the configuration vector introduced above and the elastic matrix $\mathbf{K}$ defined in Eq.~\eqref{eqKmatrix}, the scalar overdamped Eqs. (\ref{over1})-(\ref{over3}) become in matrix form
\begin{equation}\label{eqmodelprin}
 d\mathbf{r}(t) = -\mathbf{A}\mathbf{r}(t)\,dt + \sqrt{2D_0}\,d\mathbf{W}(t),
\end{equation}
where $D_0 := k_{\text{B}} T/\gamma$, $\mathbf{A}:=\mathbf{K}/\gamma$ (also called the drift matrix), and $\mathbf{W}(t)$ is an $N$-dimensional Wiener process.

For $\delta\neq 0$, one has $k_L\neq k_R$, and therefore $ A_{n,n-1}=-\frac{k_L}{\gamma}
 \neq
 -\frac{k_R}{\gamma}=A_{n-1,n}$.
Hence the drift matrix is non-symmetric and thus directly encoding the directional imbalance between two bond couplings.

\begin{figure}[h]
 \centering
  \includegraphics[width=\columnwidth]{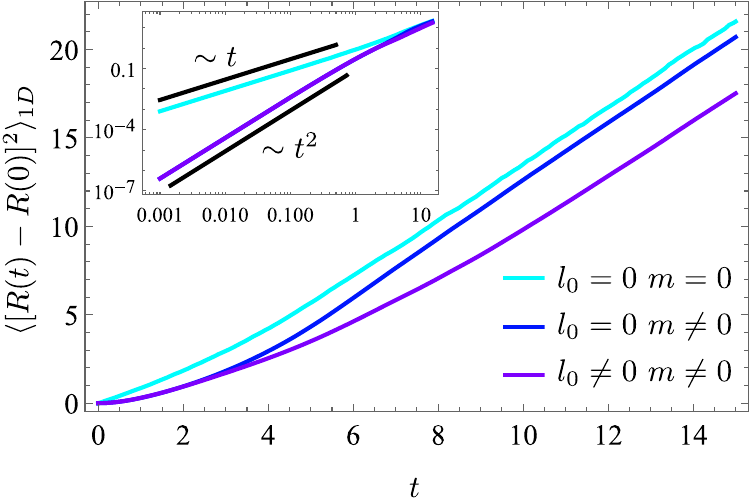}
 \caption{Numerical results for the center-of-mass MSD vs time for the three models considered in this work: underdamped nonlinear with finite natural length ($l_0>0$), underdamped linear ($l_0=0$), and overdamped linear ($l_0=0$). The transient behavior changes because of inertia and nonlinear bond relaxation, while the long-time slopes approach the same effective diffusivity. The inset shows the short-time ballistic scaling of the underdamped models and the crossover to diffusion.}
 \label{figmodelo}
\end{figure}

Let us now define the arithmetic center of mass as
\begin{equation}\label{defCMandc}
R(t):=\mathbf{c}^T\mathbf{r}(t),
\qquad
\mathbf{c}:=\frac{1}{N}(1,\dots,1)^T,
\end{equation}
which will be applied to the three studied cases: the overdamped linear, underdamped linear, and the model with finite-$l_0$ which is nonlinear. As a first step, the center of mass MSD ($R(t)$) vs time is numerically obtained for the latter three models. This is  shown in
Fig.~\ref{figmodelo}. One can see that their short- and intermediate-time behaviors are different, but at long times the curves become linear with the same slope. In Sec.~\ref{secunderdampedloti} we show analytically why the underdamped linear and the finite-$l_0$ models recover the same effective diffusivity. 

\section{Spectral structure of the drift matrix}
\label{seccispectral}
Let us first analyze the spectrum of the drift matrix since it determines both, the internal relaxation and the long-time diffusion. The algebraic details are given in Appendix~\ref{apendsym}.

Since the deterministic internal forces depend solely on relative coordinates, the drift operator is invariant under global translations. In the linear overdamped limit this gives $\mathbf A\mathbf 1=\mathbf0$, where $\mathbf 1:=(1,\dots,1)^T$. Thus $\lambda_1=0$ is an eigenvalue of $\mathbf A$, with $\mathbf1$ spanning the right nullspace associated with rigid translations. This is the familiar translational mode of the reciprocal Rouse model~\cite{DoiEdwards1986}, and here, it directly follows from the row sums of the nonreciprocal elastic matrix.

Unlike in the reciprocal Rouse chain, the arithmetic center of mass $R(t)$ (Eq.~\ref{defCMandc}) of the nonreciprocal model is not dynamically decoupled. For $\mathbf{c}^T$ to be a left zero mode, it would have to satisfy $\mathbf{c}^T\mathbf A=\mathbf0^T$ with $\mathbf A=\mathbf K/\gamma$. Computing $\mathbf{c}^T\mathbf K$ by column summation yields ($2\le j\le N-1$):
\begin{align}
(\mathbf{c}^T \mathbf{K})_1 &= \frac{1}{N}(K_{11} + K_{21}) = \frac{1}{N}(k_R - k_L), \\
(\mathbf{c}^T \mathbf{K})_j &= \frac{1}{N}(K_{j-1,j} + K_{jj} + K_{j+1,j}) = 0, \\
(\mathbf{c}^T \mathbf{K})_N &= \frac{1}{N}(K_{N-1,N} + K_{NN}) = \frac{1}{N}(k_L - k_R).
\end{align}
Consequently, for any nonreciprocal coupling ($k_L \neq k_R$),
\begin{equation}
\mathbf{c}^T \mathbf{A} = \frac{1}{\gamma N} \begin{pmatrix} k_R - k_L, & 0, & \dots, & 0, & k_L - k_R \end{pmatrix} \neq \mathbf{0}^T.
\end{equation}
Thus, $\mathbf{c}^T$ is not a left zero mode of $\mathbf{A}$. Writing $\boldsymbol{\eta}(t)=d\mathbf W(t)/dt$ formally, the projection of Eq.~\eqref{eqmodelprin} gives
\begin{equation}
 \dot R(t)
 =
 -\frac{k_R-k_L}{\gamma N}\,[r_1(t)-r_N(t)]
 +\sqrt{2D_0}\,\mathbf{c}^T\boldsymbol{\eta}(t),
\end{equation}
which shows explicitly that the arithmetic center of mass remains coupled to internal boundary deformations at finite times.

Although $\mathbf A$ is nonsymmetric when $k_L\neq k_R$, its tridiagonal structure allows it to be mapped to a real symmetric matrix~\cite{gantmacher1959theory,parlett1998symmetric,horn2012matrix}. To do so, let us define the coupling asymmetry parameter
\begin{equation}\label{defq}
q:=\frac{k_R}{k_L}=\frac{1-\delta}{1+\delta}>0.
\end{equation}
The nonsymmetric matrix $\mathbf{A}$ can then be mapped into a symmetric tridiagonal form $\mathbf{A}_s$ via the similarity transformation $\mathbf{A} = \mathbf{S} \mathbf{A}_s \mathbf{S}^{-1}$, with the positive diagonal matrix
\begin{equation}
\mathbf{S}:=\diag\!\left(1, q^{-1/2}, q^{-1}, \dots, q^{-(N-1)/2}\right).
\end{equation}
Defining $\mathbf{A}_s:=\mathbf{S}^{-1}\mathbf{A}\mathbf{S}$ yields a real symmetric matrix with diagonal entries identical to those of $\mathbf A$ and off-diagonal entries $(\mathbf A_s)_{n,n+1}=(\mathbf A_s)_{n+1,n} =-\frac{\sqrt{k_Lk_R}}{\gamma}$. Because $\mathbf{A}$ and $\mathbf{A}_s$ share the same spectrum under similarity, the spectral theorem for real symmetric matrices guarantees that all eigenvalues $\lambda_n$ of $\mathbf{A}$ are real \cite{horn2012matrix}. This rules out complex relaxation eigenmodes in the overdamped linear dynamics. Furthermore, the diagonal similarity ensures that $\mathbf{A}$ is fully diagonalizable over $\mathbb{R}$ with a complete set of linearly independent eigenvectors, ruling out nontrivial Jordan blocks.

The latter result also shows that the drift matrix has one zero eigenvalue and $N-1$ positive eigenvalues,
$0=\lambda_1<\lambda_2\le\cdots\le\lambda_N$. The zero mode corresponds to a rigid translation, whereas the remaining modes describe internal deformations. A direct proof based on the quadratic form of the symmetrized matrix is given in Appendix~\ref{apendsym}; the result is also consistent with the theory of irreducible singular $M$-matrices~\cite{berman1994nonnegative,varga2009matrix}.

\subsection{Stable subspace and semigroup decay}
Let $\boldsymbol{\psi}_1$ denote the left zero mode normalized by $\boldsymbol{\psi}_1^T\mathbf 1=1$, and define the spectral projectors $\mathbf P_0:=\mathbf 1\,\boldsymbol{\psi}_1^T$ and $\mathbf P_s:=\mathbf I-\mathbf P_0$. Then $\mathbf P_0$ projects onto the translational right-zero-mode subspace and $\mathbf P_s$ projects onto the invariant stable subspace generated by the right eigenvectors associated with $\lambda_2,\ldots,\lambda_N$. Because
\begin{equation}
 \mathbf A\mathbf 1=\mathbf 0,
 \qquad
 \boldsymbol{\psi}_1^T\mathbf A=\mathbf 0^T,
\end{equation}
one has $\mathbf A\mathbf P_s=\mathbf P_s\mathbf A=\mathbf A$.

On the stable sector, the deterministic contribution decays exponentially. Since $\mathbf A$ is diagonalizable and $\lambda_p>0$ for $p\ge2$, there exists a constant $C>0$ such that
\begin{equation}
 \left\|e^{-\mathbf A t}\mathbf P_s\right\|
 \le C e^{-\lambda_2 t},
 \qquad t\ge0.
 \label{eccstsemigroupdecay}
\end{equation}
Hence the dependence on the initial internal configuration decays exponentially. Note that, in the stochastic problem, the internal coordinates do not vanish; they relax toward bounded stationary fluctuations.

\subsection{Invariant Gaussian measure on the stable subspace}
Define the stable component of the configuration by $\mathbf r_s(t):=\mathbf P_s\mathbf r(t)$. Projecting Eq.~\eqref{eqmodelprin} gives
\begin{equation}
 d\mathbf r_s(t)
 =
 -\mathbf A\mathbf r_s(t)\,dt
 +
 \sqrt{2D_0}\,\mathbf P_s\,d\mathbf W(t).
\end{equation}
The projected noise is generally anisotropic in Euclidean coordinates because $\mathbf P_s$ is not necessarily an orthogonal projector. Its covariance is $2D_0\,\mathbf P_s\mathbf P_s^T\,dt$. The stable sector therefore admits a unique invariant Gaussian measure with second-moment matrix
\begin{equation}
 \mathbf{C}_{\infty}^{(s)}
 :=
 2D_0\int_0^\infty
 e^{-\mathbf A\tau}\mathbf P_s\mathbf P_s^T
 e^{-\mathbf A^T\tau}\,d\tau,
 \label{equaststatsecmom}
\end{equation}
where the integral converges by Eq.~\eqref{eccstsemigroupdecay}. It satisfies
\begin{equation}
 \mathbf A\mathbf{C}_{\infty}^{(s)}
 +
 \mathbf{C}_{\infty}^{(s)}\mathbf A^T
 =
 2D_0\mathbf P_s\mathbf P_s^T,
 \label{eccstalyapunov}
\end{equation}
together with $\mathbf P_s\mathbf{C}_{\infty}^{(s)}\mathbf P_s^T=\mathbf{C}_{\infty}^{(s)}$. Therefore, the internal deformation sector reaches a stationary Gaussian regime, while the full configuration vector remains nonstationary because of the translational zero mode. Standard references for Ornstein--Uhlenbeck processes and invariant Gaussian measures include Da Prato and Zabczyk, \O{}ksendal, and Gardiner~\cite{da2014stochastic,oksendal2003stochastic,gardiner2009stochastic}.

\section{Exact transient MSD for the overdamped case}
\label{seccionmsd}
We now calculate the exact transient MSD of the arithmetic center of mass for the overdamped linear model. We assume that $\mathbf r(0)$ has a finite second moment and is independent of future Wiener increments. Since the Cartesian components are independent and identically distributed, it is sufficient to work first in one spatial direction; the $d$-dimensional result follows by summing the contributions of all components.

\subsection{Differential Lyapunov equation for the second moment}
To obtain the evolution of the uncentered second-moment matrix, we start from the It\^o product rule,
\begin{equation}
 d(\mathbf{r}\mathbf{r}^T) = (d\mathbf{r})\mathbf{r}^T + \mathbf{r}(d\mathbf{r})^T + (d\mathbf{r})(d\mathbf{r})^T.
\end{equation}
Using Eq.~\eqref{eqmodelprin}, we obtain
\begin{align}
 (d\mathbf{r})\mathbf{r}^T &= -\mathbf{A}(\mathbf{r}\mathbf{r}^T)\,dt + \sqrt{2D_0}\,(d\mathbf{W})\mathbf{r}^T,
 \\
 \mathbf{r}(d\mathbf{r})^T &= -(\mathbf{r}\mathbf{r}^T)\mathbf{A}^T\,dt + \sqrt{2D_0}\,\mathbf{r}\,d\mathbf{W}^T,
 \\
 (d\mathbf{r})(d\mathbf{r})^T &= 2D_0\,\mathbf{I}\,dt.
\end{align}
Therefore,
\begin{equation}
    \begin{split}
        d(\mathbf{r}\mathbf{r}^T)
 &=
 \left[-\mathbf{A}(\mathbf{r}\mathbf{r}^T)-(\mathbf{r}\mathbf{r}^T)\mathbf{A}^T+2D_0\mathbf{I}\right]dt \\
             &\quad + \sqrt{2D_0}\left[(d\mathbf{W})\mathbf{r}^T+\mathbf{r}\,d\mathbf{W}^T\right].
    \end{split}
\end{equation}
Taking the ensemble average, the stochastic terms vanish and we define the uncentered second-moment matrix
\begin{equation}
 \mathbf{C}(t):=\langle \mathbf{r}(t)\mathbf{r}(t)^T\rangle.
\end{equation}
It coincides with the covariance matrix only when $\langle\mathbf r(t)\rangle=\mathbf 0$. With this notation, the second moment obeys the differential Lyapunov equation
\begin{equation}
 \dot{\mathbf{C}}(t) = -\mathbf{A}\mathbf{C}(t)-\mathbf{C}(t)\mathbf{A}^T + 2D_0\mathbf{I}.
 \label{ecuaLyap}
\end{equation}
Its solution is given by
\begin{equation}
 \mathbf{C}(t)=e^{-\mathbf{A}t}\mathbf{C}_0 e^{-\mathbf{A}^T t}
 +2D_0\int_0^t e^{-\mathbf{A}\tau}e^{-\mathbf{A}^T\tau}\,d\tau,
 \label{eccCtexacto}
\end{equation}
where we have defined the initial second moment as
\begin{equation}\label{ecuaC0}
 \mathbf{C}_0:=\mathbf{C}(0)=\langle \mathbf{r}(0)\mathbf{r}(0)^T\rangle.
\end{equation}
A short derivation is included in Appendix~\ref{apelyap}.

The OU solution can be written as
\begin{equation}
 \mathbf{r}(t)=e^{-\mathbf{A}t}\mathbf{r}(0)+\sqrt{2D_0}\int_0^t e^{-\mathbf{A}(t-s)}\,d\mathbf{W}(s).
\end{equation}
Multiplying by $\mathbf{r}(0)^T$ and taking expectation yields
\begin{equation}
 \langle \mathbf{r}(t)\mathbf{r}(0)^T\rangle = e^{-\mathbf{A}t}\mathbf{C}_0,
 \label{ecccrosscov}
\end{equation}
since Wiener increments for $s>0$ are independent of the initial condition.

\subsection{Exact MSD formula}
The averaged center of mass in one-dimension is
\begin{equation}
 \langle [R(t)-R(0)]^2\rangle_{1D}
 =
 \langle R(t)^2\rangle + \langle R(0)^2\rangle -2\langle R(t)R(0)\rangle.
\end{equation}
Notice that each term can be expressed through the one- and two-time second moments, namely
\begin{align}
 \langle R(t)^2\rangle &= \mathbf{c}^T\mathbf{C}(t)\mathbf{c},
 \\
 \langle R(0)^2\rangle &= \mathbf{c}^T\mathbf{C}_0\mathbf{c},
 \\
 \langle R(t)R(0)\rangle &= \mathbf{c}^T e^{-\mathbf{A}t}\mathbf{C}_0\mathbf{c}.
\end{align}
Thus by substituting $\mathbf{C}(t)$ from Eq.~\eqref{eccCtexacto}, one obtains
\begin{align}\label{eccMSDexacto}
\langle [R(t)-R(0)]^2\rangle_{1D} \nonumber
 =\mathbf{c}^T(e^{-\mathbf{A}t}-\mathbf{I})\mathbf{C}_0(e^{-\mathbf{A}^T t}-\mathbf{I})\mathbf{c}\\ 
 +2D_0\,\mathbf{c}^T\left(\int_0^t e^{-\mathbf{A}\tau}e^{-\mathbf{A}^T\tau}\,d\tau\right)\mathbf{c},
\end{align}
which is the exact transient MSD of the center of mass in one Cartesian component. Equation~\eqref{eccMSDexacto} contains a transient  MSD term, determined by the initial second moment $\mathbf{C}_0$, and a noise-induced term, determined by the OU kernel $e^{-\mathbf A\tau}e^{-\mathbf A^T\tau}$. The latter contains the long-time diffusive contribution generated by the translational zero mode. 

Finally, for $d$ independent and identically distributed Cartesian components, the full MSD is
\begin{equation}
 \left\langle \|\mathbf R(t)-\mathbf R(0)\|^2\right\rangle
 =
 d\,\left\langle [R(t)-R(0)]^2\right\rangle_{1D},
\end{equation}
where $\mathbf R(t)$ denotes the $d$-dimensional arithmetic center-of-mass vector. The next section explicitly calculates the system's MSD by applying a mode expansion technique.

\section{Mode expansion and long-time effective diffusivity}
\label{seccmode}
We now use the biorthogonal spectral decomposition of the nonsymmetric drift matrix to rewrite Eq.~\eqref{eccMSDexacto}. Since $\mathbf{A}$ is diagonalizable, there exist right and left eigenvectors $\boldsymbol{\phi}_p$ and $\boldsymbol{\psi}_p$ such that
\begin{equation}
 \mathbf{A}\boldsymbol{\phi}_p=\lambda_p\boldsymbol{\phi}_p,
 \qquad
 \boldsymbol{\psi}_p^T\mathbf{A}=\lambda_p\boldsymbol{\psi}_p^T,
 \qquad
 \boldsymbol{\psi}_p^T\boldsymbol{\phi}_q=\delta_{pq}.
\end{equation}
For the translational mode we choose $\boldsymbol{\phi}_1:=\mathbf 1$, consistently with $\boldsymbol{\psi}_1^T\mathbf1=1$; hence $\mathbf{c}^T\boldsymbol{\phi}_1=1$.
Thus,
\begin{equation}
 \mathbf{I}=\sum_{p=1}^N \boldsymbol{\phi}_p\boldsymbol{\psi}_p^T,
 \qquad
 e^{-\mathbf{A}t}=\sum_{p=1}^N e^{-\lambda_p t}\boldsymbol{\phi}_p\boldsymbol{\psi}_p^T.
\end{equation}
We now define the projections relevant to the arithmetic center of mass and the initial second moment by
\begin{equation}\label{eccmode_projections}
 a_p:=\mathbf{c}^T\boldsymbol{\phi}_p,
 \qquad
 C_{pq}^{(0)}:=\boldsymbol{\psi}_p^T\mathbf{C}_0\boldsymbol{\psi}_q,
 \qquad
 G_{pq}:=\boldsymbol{\psi}_p^T\boldsymbol{\psi}_q.
\end{equation}
Then the exact MSD becomes
\begin{align}
\langle [R(t)-R(0)]^2\rangle_{1D}
 &=
 \sum_{p,q=1}^N
 a_p a_q C_{pq}^{(0)}
 (e^{-\lambda_p t}-1)
 \nonumber\\
 &\qquad\times
 (e^{-\lambda_q t}-1)
 \nonumber\\
 &\quad
 +2D_0\sum_{p,q=1}^N
 a_p a_q G_{pq}
 \nonumber\\
 &\qquad\times
 \int_0^t
 e^{-(\lambda_p+\lambda_q)\tau}\,d\tau.
\end{align}
Evaluating the integral and separating the $(p,q)=(1,1)$ contribution yields
\begin{align}\label{eccMSDmode}
 \langle [R(t)-R(0)]^2\rangle_{1D}
 &=
 \sum_{p,q=1}^N
 a_p a_q C_{pq}^{(0)}
 (e^{-\lambda_p t}-1)
 \nonumber\\
 &\qquad\times
 (e^{-\lambda_q t}-1)
 +2D_0 a_1^2G_{11}t
 \nonumber\\
 &\quad
 +2D_0
 \sum_{(p,q)\neq(1,1)}
 a_p a_q G_{pq}
 \nonumber\\
 &\qquad\times
 \frac{1-e^{-(\lambda_p+\lambda_q)t}}
 {\lambda_p+\lambda_q}.
\end{align}
Notice that the latter expression is valid for any time. Actually, it will be numerically validated in Section \ref{secnum}.
Additionally, one can notice that the only term that grows linearly in time is the zero-mode contribution,
\begin{align}\label{msdlinearterm}
2D_0a_1^2G_{11}t
&=
2D_0
\left(\mathbf{c}^T\boldsymbol{\phi}_1\right)^2
\boldsymbol{\psi}_1^T\boldsymbol{\psi}_1\,t
\nonumber\\
&=
2D_0\boldsymbol{\psi}_1^T\boldsymbol{\psi}_1\,t.
\end{align}

To identify the diffusive contribution, let us project Eq.~\eqref{eqmodelprin} onto the left zero mode:
\begin{equation}\label{eccX1zeromodo}
 X_1(t):=\boldsymbol{\psi}_1^T\mathbf{r}(t).
\end{equation}
Since $\boldsymbol{\psi}_1^T\mathbf{A}=\mathbf{0}^T$,
\begin{equation}
 dX_1(t)=\sqrt{2D_0}\,\boldsymbol{\psi}_1^T d\mathbf{W}(t).
\end{equation}
Hence
\begin{equation}
 \langle [X_1(t)-X_1(0)]^2\rangle = 2D_0(\boldsymbol{\psi}_1^T\boldsymbol{\psi}_1)t.
\end{equation}
This term is exactly the zero-mode contribution in Eq.~\eqref{msdlinearterm}, so the long-time linear growth of the arithmetic center-of-mass MSD is
\begin{equation}
 \langle [R(t)-R(0)]^2\rangle_{1D} \sim 2D_\eff t,
 \qquad
 D_\eff := D_0\,\boldsymbol{\psi}_1^T\boldsymbol{\psi}_1.
 \label{eccDeffgen}
\end{equation}
Therefore, the long-time diffusivity is determined by the left zero mode rather than by the arithmetic averaging vector $\mathbf c$.

For the nonreciprocal tridiagonal chain, the left zero mode is determined by $\mathbf{A}^T\boldsymbol{\psi}_1=\mathbf{0}$. Writing $\boldsymbol{\psi}_1=(\psi_{1,1},\dots,\psi_{1,N})^T$, the interior recurrence for $N\ge3$ and $2\le n\le N-1$ is
\begin{equation}\label{eqrecc}
 -k_L\psi_{1,n+1} + (k_L+k_R)\psi_{1,n} - k_R\psi_{1,n-1}=0.
\end{equation}
Finally, by using Eq.~\eqref{defq}, the boundary conditions, the recurrence relation~\eqref{eqrecc} when $N\ge3$, and the biorthogonal normalization $\boldsymbol{\psi}_1^T\boldsymbol{\phi}_1=1$, one obtains the result below (see Appendix~\ref{appleftzero}); the same formula follows directly from the boundary equations when $N=2$:
\begin{equation}
 \psi_{1,n} = \frac{q^{n-1}}{\sum_{m=0}^{N-1} q^m},
 \qquad n=1,\dots,N.
 \label{eccizqzeroexplicito}
\end{equation}
Then, by substituting Eq.~\eqref{eccizqzeroexplicito} into Eq.~\eqref{eccDeffgen} leads to
\begin{equation}
 D_\eff(q,N) = D_0\,
 \frac{\sum_{n=0}^{N-1} q^{2n}}{\left(\sum_{n=0}^{N-1} q^n\right)^2}.
\end{equation}
For $q\neq 1$, this becomes according to Appendix \ref{appleftzero}
\begin{equation}
 D_\eff(q,N)
 = D_0\,
 \frac{(1-q)(1+q^N)}{(1+q)(1-q^N)}.
 \label{eccDeffcerrado}
\end{equation}
which is actually the new effective diffusion  of a non-reciprocal overdamped Rouse chain freely diffusing.
Note that in the reciprocal limit $q\to 1$, one recovers $D_\eff \to D_0 /N$, which is the standard result of a reciprocal Rouse chain.
Equation~\eqref{eccDeffcerrado} is one of the main results of this work. It explicitly shows how non-reciprocity changes the long-time center-of-mass diffusivity through the nonuniform left zero mode. Basically, Eq.~\eqref{eccDeffcerrado} indicates that diffusion is enhanced as nonreciprocity in the chain grows. This analytical result will be numerically validated in Section \ref{secnum}.
\section{Connection between a non-reciprocal chain and an active chain tangentially driven}
Equation (\ref{eccDeffcerrado}) represents the effective diffusion coefficient of a  discrete non-reciprocal chain. Interestingly,  by  defining  $q=\exp(-2\alpha/N)$, Eq. (\ref{eccDeffcerrado}) becomes $D_0\tanh(\alpha/N)\coth(\alpha)$, which in the limit $N\to \infty$  gives
\begin{equation}
 D_\eff \sim
 \frac{D_0}{N}\alpha \coth(\alpha).
 \label{limit}
\end{equation}
The above result is precisely the expression obtained for a continuum active Rouse chain reported in \cite{Peterson}. This proves that  our result is general, that is, it is valid for both, the discrete and the continuum cases of an active Rouse chain. Moreover, this results relates two seemingly different problems, namely, an active Rouse chain tangentially driven with a non-reciprocal Rouse chain. Therefore, one can think of as adding tangentially active forces to an elastic chain as a means to generate non-reciprocity in the system. Conversely, one can think of  a non-reciprocal chain as an active chain.

\section{Long-time diffusivity from the translational coordinate}
\label{sectzeromodedom}
The previous  linear analysis identifies a weighted translational coordinate directly related to the MSD. We now formally establish the connection between this coordinate and the arithmetic center of mass diffusivity. The only condition needed is that the remaining contribution has a displacement with uniformly bounded second moment.

A closely related argument is used in transport theory, where bounded non-diffusive terms do not modify transport coefficients~\cite{GrasselliBaroni2021}. For completeness, we state and prove the form needed here.

\begin{proposition}[Long time diffusivity transport coefficient from the translational coordinate]
\label{thm:zero_mode_dominance}
Let $\mathbf R(t),\mathbf X(t),\mathbf Y(t)\in\mathbb R^d$ be stochastic processes with finite second moments for every $t\ge0$, satisfying
\begin{equation}
 \mathbf R(t)=\mathbf X(t)+\mathbf Y(t).
 \label{ecctransfer_decomposition}
\end{equation}
Assume that
\begin{equation}
 \lim_{t\to\infty}
 \frac{\left\langle\|\mathbf X(t)-\mathbf X(0)\|^2\right\rangle}{2d\,t}
 =
 D_{\eff}\in[0,\infty),
 \label{ecctransfer_X}
\end{equation}
and that the displacement of the remainder is uniformly bounded in its second moment,
\begin{equation}
 \sup_{t\ge0}
 \left\langle
 \|\mathbf Y(t)-\mathbf Y(0)\|^2
 \right\rangle
 <\infty.
 \label{ecctransfer_Y}
\end{equation}
Then
\begin{equation}
 \lim_{t\to\infty}
 \frac{\left\langle\|\mathbf R(t)-\mathbf R(0)\|^2\right\rangle}{2d\,t}
 =
 D_{\eff}.
 \label{ecctransfer_result}
\end{equation}
\end{proposition}

\begin{proof}
Set $\Delta\mathbf R(t):=\mathbf R(t)-\mathbf R(0)$, $\Delta\mathbf X(t):=\mathbf X(t)-\mathbf X(0)$, and $\Delta\mathbf Y(t):=\mathbf Y(t)-\mathbf Y(0)$. From Eq.~\eqref{ecctransfer_decomposition}, $\Delta\mathbf R(t)=\Delta\mathbf X(t)+\Delta\mathbf Y(t)$. Therefore,
\begin{align}
 \left\langle\|\Delta\mathbf R(t)\|^2\right\rangle
 &=
 \left\langle\|\Delta\mathbf X(t)\|^2\right\rangle
 +
 \left\langle\|\Delta\mathbf Y(t)\|^2\right\rangle
 \nonumber\\
 &\quad
 +
 2\left\langle
 \Delta\mathbf X(t)\cdot\Delta\mathbf Y(t)
 \right\rangle.
 \label{ecctransfer_expand}
\end{align}
By assumption, the second term is $O(1)$. For the cross term, Cauchy--Schwarz gives
\begin{equation}
 \left|
 \left\langle
 \Delta\mathbf X(t)\cdot\Delta\mathbf Y(t)
 \right\rangle
 \right|
 \le
 \left\langle\|\Delta\mathbf X(t)\|^2\right\rangle^{1/2}
 \left\langle\|\Delta\mathbf Y(t)\|^2\right\rangle^{1/2}.
\end{equation}
Equation~\eqref{ecctransfer_X} implies
$\left\langle\|\Delta\mathbf X(t)\|^2\right\rangle=O(t)$, while Eq.~\eqref{ecctransfer_Y} implies
$\left\langle\|\Delta\mathbf Y(t)\|^2\right\rangle=O(1)$. Hence the cross term is $O(t^{1/2})$. Dividing Eq.~\eqref{ecctransfer_expand} by $2dt$ and taking $t\to\infty$ proves Eq.~\eqref{ecctransfer_result}.
\end{proof}

Let us  exemplify the above result for a linear model. 
For one Cartesian component, the right and left eigenvectors $\boldsymbol{\phi}_p$ and $\boldsymbol{\psi}_p$ give
\begin{equation}\label{eccmode_coordinates}
 \mathbf r(t)
 =
 \sum_{p=1}^{N}
 X_p(t)\boldsymbol{\phi}_p,
 \qquad
 X_p(t):=\boldsymbol{\psi}_p^T\mathbf r(t),
\end{equation}
where $X_p(t)$ are the position amplitudes, and because $\mathbf{c}^T\boldsymbol{\phi}_1=1$,
\begin{equation}
 R(t)
 =
 X_1(t)
 +
 \sum_{p=2}^{N}a_pX_p(t),
 \label{ecclinear_transfer_decomposition}
\end{equation}
where $a_p$ is defined in Eq.~\eqref{eccmode_projections}.
From the above expressions, one may identify $X=X_1$ and
$Y=\sum_{p=2}^{N}a_pX_p$.  Now, if the internal-mode displacements are uniformly bounded in their second moments, then the remainder in Eq.~\eqref{ecclinear_transfer_decomposition} satisfies Eq.~\eqref{ecctransfer_Y}, hence Proposition~\ref{thm:zero_mode_dominance} shows that  coordinate $X_1$ provides the long-time diffusion coefficient  of the arithmetic center of mass.

Note that the above result does not require a linear modal decomposition. For the finite-$l_0$ nonlinear chain we will use instead the exact relation $\mathbf R-\mathbf X=\sum_j A_j\mathbf b_j$. In that case, bounded bond fluctuations play the same role as the stable internal modes of the linear chain.

\section{Long-time MSD of the underdamped models}
\label{secunderdampedloti}

We now turn to the underdamped models. Although the exact transient result of Sec.~\ref{seccionmsd} applies to the overdamped linear chain, the same long-time diffusivity can also be obtained for the underdamped linear and nonlinear models. To show this, we use Proposition~\ref{thm:zero_mode_dominance}. For the regularized interaction force giving rise to the nonlinear Rouse chain, its required bond confinement is proved in Appendix~\ref{appconfinement}.

\subsection{Underdamped linear chain with $l_0=0$}

For a generic Cartesian component, the underdamped linear model is
\begin{align}
 d\mathbf r(t) &= \mathbf v(t)\,dt,
 \\
 d\mathbf v(t) &=
 -\frac{1}{m}\mathbf K\mathbf r(t)\,dt
 -\frac{\gamma}{m}\mathbf v(t)\,dt
 +\frac{\sqrt{2\gamma k_{\text{B}} T}}{m}\,d\mathbf W(t).
 \label{eccunderdamped_linear_sde}
\end{align}
Equivalently, with $\mathbf z(t)=(\mathbf r(t),\mathbf v(t))^T$,
\begin{equation}
 d\mathbf z(t)=\mathbf B\mathbf z(t)\,dt+\mathbf G\,d\mathbf W(t),
\end{equation}
where
\begin{equation}
 \mathbf B:=
 \begin{pmatrix}
  \mathbf 0 & \mathbf I\\
  -\mathbf K/m & -(\gamma/m)\mathbf I
 \end{pmatrix},
 \qquad
 \mathbf G:=
 \begin{pmatrix}
  \mathbf 0\\
  \sqrt{2\gamma k_{\text{B}} T}\,\mathbf I/m
 \end{pmatrix}.
\end{equation}
Thus the underdamped linear chain is an Ornstein--Uhlenbeck process in phase space.

Let $\kappa_p:=\gamma\lambda_p$ denote the eigenvalues of $\mathbf K$, so that $\kappa_1=0$ and $\kappa_p>0$ for $p\ge2$. Using the position amplitudes $X_p(t)$ defined in Eq.~\eqref{eccmode_coordinates}, one defines the velocity amplitudes by $U_p(t):=\boldsymbol{\psi}_p^T\mathbf v(t)$.
Projection onto the $p$th left eigenvector gives
\begin{align}
 dX_p(t) &= U_p(t)\,dt,
 \\
 m\,dU_p(t)
 &=
 -\kappa_p X_p(t)\,dt
 -\gamma U_p(t)\,dt
 \nonumber\\
 &\quad
 +\sqrt{2\gamma k_{\text{B}} T}\,
 \boldsymbol{\psi}_p^T d\mathbf W(t).
 \label{eccunderdamped_mode_equations}
\end{align}
For the translational mode, $\kappa_1=0$. Writing
\begin{equation}
 X(t):=X_1(t)=\boldsymbol{\psi}_1^T\mathbf r(t),
 \qquad
 U(t):=U_1(t)=\boldsymbol{\psi}_1^T\mathbf v(t),
\end{equation}
one obtains
\begin{align}
 dX(t) &= U(t)\,dt,
 \\
 m\,dU(t) &= -\gamma U(t)\,dt
 +\sqrt{2\gamma k_{\text{B}} T}\,\boldsymbol{\psi}_1^T d\mathbf W(t).
 \label{eccprojected_underdamped}
\end{align}
Set $g_1:=\boldsymbol{\psi}_1^T\boldsymbol{\psi}_1$ and $\tau_m:=\frac{m}{\gamma}$. For an initial projected velocity $U(0)$ with finite second moment, independent of future Wiener increments, a direct integration of Eq.~\eqref{eccprojected_underdamped} yields
\begin{align}
 \left\langle [X(t)-X(0)]^2\right\rangle
 &=
 \tau_m^2\left(1-e^{-t/\tau_m}\right)^2
 \left\langle U(0)^2\right\rangle
 \nonumber\\
 &\quad
 +2D_0g_1
 \Bigg[
 t
 -2\tau_m\left(1-e^{-t/\tau_m}\right)
 \nonumber\\
 &\hspace{4.3em}
 +\frac{\tau_m}{2}
 \left(1-e^{-2t/\tau_m}\right)
 \Bigg].
 \label{eccunderdamped_projected_msd_general}
\end{align}
If the projected velocity is initially in its stationary Gaussian distribution,
\begin{equation}
 \left\langle U(0)^2\right\rangle
 =
 \frac{k_{\text{B}}T}{m}g_1
 =
 \frac{D_0g_1}{\tau_m},
\end{equation}
Eq.~\eqref{eccunderdamped_projected_msd_general} reduces to
\begin{equation}
 \left\langle [X(t)-X(0)]^2\right\rangle
 =
 2D_0g_1
 \left[
 t-\tau_m\left(1-e^{-t/\tau_m}\right)
 \right].
 \label{eccunderdamped_projected_msd}
\end{equation}
For arbitrary finite-second-moment initial velocities, the difference from Eq.~\eqref{eccunderdamped_projected_msd} is bounded as $t\to\infty$. Therefore,
\begin{equation}
 \left\langle [X(t)-X(0)]^2\right\rangle
 =
 2D_{\eff}t+O(1),
 \qquad
 D_{\eff}=D_0g_1.
 \label{eccX_long_time}
\end{equation}
Let us now verify the bounded-remainder condition required by Proposition~\ref{thm:zero_mode_dominance}. For each internal mode $p\ge2$, the deterministic part of Eq.~\eqref{eccunderdamped_mode_equations} has characteristic exponents
\begin{equation}
 \mu_{p,\pm}
 =
 \frac{-\gamma\pm\sqrt{\gamma^2-4m\kappa_p}}{2m}.
 \label{eccinternal_phase_eigenvalues}
\end{equation}
Because $\gamma>0$, $m>0$, and $\kappa_p>0$, we have $\operatorname{Re}\mu_{p,\pm}<0$. Hence every internal phase-space mode is exponentially stable. Under initial conditions with finite second moments, the associated stochastic oscillator has uniformly bounded position and velocity second moments. In particular,
\begin{equation}
 \sup_{t\ge0}
 \left\langle [X_p(t)-X_p(0)]^2\right\rangle
 <\infty,
 \qquad p=2,\ldots,N.
\end{equation}
Using Eq.~\eqref{ecclinear_transfer_decomposition}, the difference
\begin{equation}
 Y(t):=R(t)-X(t)
 =
 \sum_{p=2}^{N}a_pX_p(t),
\end{equation}
has therefore a displacement with uniformly bounded second moment. Proposition~\ref{thm:zero_mode_dominance} then gives
\begin{equation}
 \lim_{t\to\infty}
 \frac{\left\langle [R(t)-R(0)]^2\right\rangle}{t}
 =
 2D_{\eff},
\end{equation}
for one Cartesian component. In $d$ independent Cartesian dimensions, one simply has
\begin{equation}
 \left\langle \|\mathbf R(t)-\mathbf R(0)\|^2\right\rangle
 =
 2dD_{\eff}t+O(t).
 \label{eccR_linear_inertial_longtime}
\end{equation}
Therefore, inertia changes the short- and intermediate-time MSD through the velocity-relaxation time $\tau_m$, but the long-time diffusive slope remains unchanged.

\subsection{Underdamped nonlinear chain with finite $l_0$}

For $l_0>0$, the interacting force in Eq.~\eqref{eqcentralf1} is nonlinear in the bond norm. It implies that the complete phase-space dynamics is no longer an Ornstein--Uhlenbeck process and the modal decomposition used above cannot be applied globally. Nevertheless, the weighted translational coordinate can still be obtained exactly. This is shown in the following lines.

For every nonzero bond vector $\mathbf b$, Eqs.~\eqref{eqcentralf1} and~\eqref{defq} imply $\mathbf f_R(\mathbf b)=q\,\mathbf f_L(\mathbf b)$. The normalized weight in Eq.~\eqref{eccizqzeroexplicito} satisfy $\psi_{1,n+1}=q\psi_{1,n}$ and $\sum_{n=1}^{N}\psi_{1,n}=1$. Now, define
\begin{equation}
 \mathbf X(t):=\sum_{n=1}^{N}\psi_{1,n}\mathbf r_n(t),
 \qquad
 \mathbf U(t):=\sum_{n=1}^{N}\psi_{1,n}\mathbf v_n(t),
\end{equation}
hence $d\mathbf X(t)=\mathbf U(t)\,dt$. By adding the nonlinear internal forces with the same weights gives
\begin{equation}
\sum_{n=1}^{N}\psi_{1,n}\mathbf F_n
 =
 \sum_{j=1}^{N-1}
 \left[
 \psi_{1,j}\mathbf f_R(\mathbf b_j)
 -
 \psi_{1,j+1}\mathbf f_L(\mathbf b_j)
 \right] = \mathbf 0 .
\label{eccnonlinear_projected_force_cancel}
\end{equation}
The cancellation is exact and does not rely on a small-deformation expansion about $b_j=l_0$. Consequently,
\begin{align}
 d\mathbf X(t) &= \mathbf U(t)\,dt,
 \\
 m\,d\mathbf U(t)
 &=
 -\gamma\mathbf U(t)\,dt
 +\sqrt{2\gamma k_{\text{B}} T}
 \sum_{n=1}^{N}\psi_{1,n}\,d\mathbf W_n(t).
 \label{eccnonlinear_projected_dynamics}
\end{align}
On the other hand, the weighted noise has covariance
\begin{equation}
 \left\langle
 \left(\sum_n\psi_{1,n}\,d\mathbf W_n\right)
 \left(\sum_m\psi_{1,m}\,d\mathbf W_m\right)^T
 \right\rangle
 =
 g_1\mathbf I_d\,dt.
\end{equation}
Thus each Cartesian component of $\mathbf X$ obeys exactly the same free inertial Langevin equation as the linear translational mode. For stationary initial projected velocities one has
\begin{equation}
 \left\langle \|\mathbf X(t)-\mathbf X(0)\|^2\right\rangle
 =
 2dD_0g_1
 \left[
 t-\tau_m\left(1-e^{-t/\tau_m}\right)
 \right],
 \label{eccX_nonlinear_msd}
\end{equation}
whereas arbitrary finite-second-moment initial projected velocities add only a bounded $O(1)$ contribution at long times. Hence
\begin{equation}
 \left\langle \|\mathbf X(t)-\mathbf X(0)\|^2\right\rangle
 =
 2dD_{\eff}t+O(1).
 \label{eccX_nonlinear_longtime}
\end{equation}
To connect $\mathbf X$ with the arithmetic center of mass $\mathbf R(t):=\frac{1}{N}\sum_{n=1}^{N}\mathbf r_n(t)$ one writes
\begin{equation}
 \mathbf r_n(t)
 =
 \mathbf r_1(t)+\sum_{j=1}^{n-1}\mathbf b_j(t).
\end{equation}
A direct rearrangement gives
\begin{equation}
 \mathbf R(t)-\mathbf X(t)
 =
 \sum_{j=1}^{N-1}A_j\mathbf b_j(t),\quad A_j:=
 \frac{N-j}{N}
 -
 \sum_{n=j+1}^{N}\psi_{1,n}.
 \label{eccR_minus_X_bonds}
\end{equation}
For $q\neq1$, this coefficient may be written explicitly as
\begin{equation}
 A_j
 =
 \frac{N-j}{N}
 -
 \frac{q^j-q^N}{1-q^N}.
 \label{eccAj_explicit}
\end{equation}
Notice that in the reciprocal limit $q\to1$, one has $\psi_{1,n}\to1/N$ and therefore $A_j\to0$, so that $\mathbf X=\mathbf R$.

For the nonreciprocal nonlinear chain, define $\mathbf Y(t):=\mathbf R(t)-\mathbf X(t)$. For the smooth regularization in Eq.~\eqref{eqregularizedforce}, the internal bonds are uniformly confined in  its second moment. More precisely, for every initial condition with finite internal energy, Appendix~\ref{appconfinement} proves
\begin{equation}
 \sup_{t\ge0}
 \left\langle \|\mathbf b_j(t)\|^2\right\rangle
 <\infty,
 \qquad
 j=1,\ldots,N-1.
 \label{eccbond_confinement}
\end{equation}
Since $\mathbf Y$ is a finite linear combination of the bond vectors, Eq.~\eqref{eccbond_confinement} implies
\begin{equation}
 \sup_{t\ge0}
 \left\langle
 \|\mathbf Y(t)-\mathbf Y(0)\|^2
 \right\rangle
 <\infty.
\end{equation}
Equations~\eqref{eccX_nonlinear_longtime} and~\eqref{eccR_minus_X_bonds} therefore satisfy the hypotheses of Proposition~\ref{thm:zero_mode_dominance}. Consequently,
\begin{equation}
 \lim_{t\to\infty}
 \frac{\left\langle \|\mathbf R(t)-\mathbf R(0)\|^2\right\rangle}{2d\,t}
 =
 D_{\eff}
 =
 D_0\boldsymbol{\psi}_1^T\boldsymbol{\psi}_1 .
 \label{eccR_nonlinear_longtime}
\end{equation}
For the singular force law in Eq.~\eqref{eqcentralf1}, the same Lyapunov estimate can be used along a global collision-free trajectory. However, this estimate does not prove collision avoidance. Therefore, the non-regularized result remains conditional on global collision-free evolution, whereas the regularized dynamics directly satisfies Eq.~\eqref{eccbond_confinement}.

Note that $l_0$ does not disappear from the long-time diffusivity because the nonlinear force becomes negligible. Instead, it cancels exactly from the weighted translational equation through Eq.~\eqref{eccnonlinear_projected_force_cancel}. The natural bond length still affects the internal relaxation and, consequently, the transient MSD.

\section{Numerical validation}
\label{secnum}
We finally compare the analytical results with numerical simulations. The exact transient expression is tested for the overdamped linear model, while the three models are compared at long times to verify their common asymptotic diffusivity (see Fig. \ref{figmodelo}).

For each stochastic trajectory, the arithmetic center-of-mass MSD is estimated as
\begin{equation}
 \MSD_\CM(t)
 :=
 \left\langle
 \|\mathbf R(t)-\mathbf R(0)\|^2
 \right\rangle.
\end{equation}
The overdamped dynamics were integrated using the Euler--Maruyama method, while the underdamped models were simulated with a Langevin operator-splitting scheme consisting of a half force kick, an exact Ornstein--Uhlenbeck velocity update, a full position drift, and a final half force kick. For the finite-$l_0$ model, the smooth regularization in Eq.~\eqref{eqregularizedforce} preserves the proportionality $\mathbf f_R^{(\varepsilon)}=q\,\mathbf f_L^{(\varepsilon)}$. Unless otherwise stated, the simulations use $N=5$, $d=2$, $k=\gamma=m=k_{\mathrm B}T=1$, $\delta=0.4$, time step $\Delta t=10^{-3}$, final time $t=15$, and $4000$ independent trajectories. Initial bead positions are sampled independently from a zero-mean Gaussian distribution with standard deviation $0.1$. In the underdamped models, the initial velocity components are sampled independently from a zero-mean Gaussian distribution with variance $k_{\mathrm B}T/m$. 
\begin{figure}[h]
 \centering
 \includegraphics[width=\columnwidth]{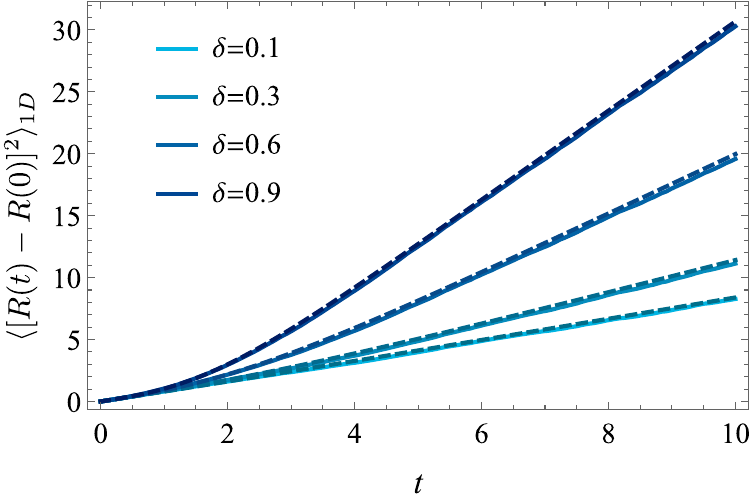}
 \caption{Comparison between the exact transient result and Brownian dynamics simulations for the MSD vs time of an overdamped linear chain as a function of $\delta$. The numerical MSDs are shown by solid lines, while the dashed lines correspond to Eq.~\eqref{eccMSDmode} and  Eq.~(\ref{eccDeffcerrado}). The long-time slopes approach the predicted values of $D_{\eff}$.}
 \label{figmsddeltacomp}
\end{figure}
The first numerical simulation is shown in Fig.~\ref{figmsddeltacomp}, where the overdamped MSD vs time and for several values of $\delta$ is plotted. Here,  the solid curves represent Brownian dynamics simulations, whereas the dashed curves represent the analytical result taken from Eq.~\eqref{eccMSDmode} and  Eq.~(\ref{eccDeffcerrado}) using the same initial second moment. One can see that numerics and theory coincide over the transient and long-time regimes. Notice how their long-time slopes approach the corresponding values of $D_{\eff}$. Therefore, the latter results validate our analytical approach.

\begin{figure}[h]
 \centering
 \includegraphics[width=\columnwidth]{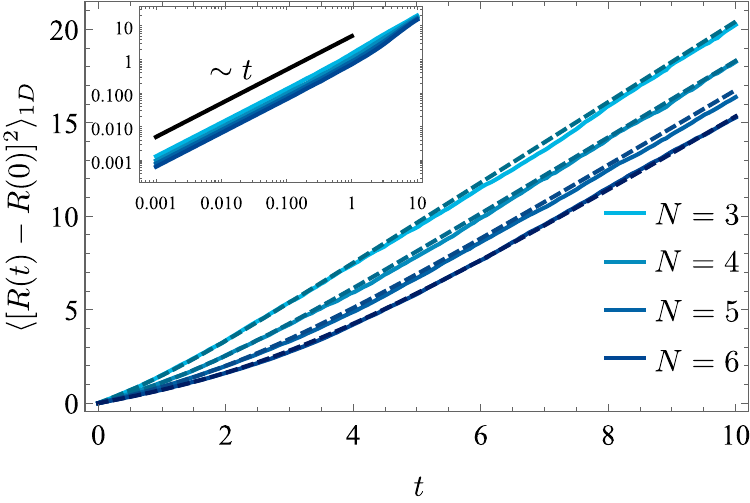}
 \caption{Comparison between the exact transient result and Brownian dynamics simulations for the MSD vs time of an overdamped nonreciprocal chain with $q=k_R/k_L=0.5$ and several values of $N$. The asymptotic slopes follow the chain-length dependence predicted by Eq.~\eqref{eccDeffcerrado} and depicted in black dashed-lines. The transient part follows from the combination of Eq.~\eqref{eccMSDmode} and  Eq.~(\ref{eccDeffcerrado}).}
 \label{figumsdNcomp}
\end{figure}

The effect of chain length on the MSD versus time diagram is shown in Fig.~\ref{figumsdNcomp}. Here $q=k_R/k_L=0.5$ and the overdamped scenario is considered. Once again, the numerical results are shown  by solid lines, while  the theoretical ones (Eq.~\eqref{eccMSDmode} and  Eq.~(\ref{eccDeffcerrado})) are shown in dashed lines. One can see that theory and simulations coincide, thus validating again our analysis. As in the Rouse model, this plot indicates that the longer the chain, the smaller its diffusivity. The inset shows its linear behavior at short times due to the overdamped scenario in consideration.

\begin{figure}[h]
 \centering
 \includegraphics[width=\columnwidth]{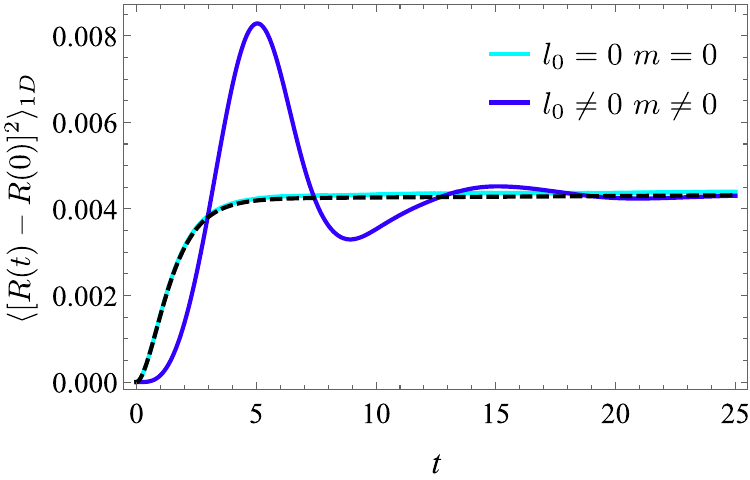}
 \caption{Mean-square displacement versus time for small noise intensities ($k_B T =  0.00001$)  and for both the over-damped and underdamped models. The analytical over-damped result, Eq. (\ref{eccMSDmode}), is plotted in dashed lines. Here, $q=k_R/k_L=0.5$. }
 \label{Fin}
\end{figure}

To close this section the dynamics of the MSD vs time for the overdamped and  underdamped models, at small thermal fluctuations ($k_B T =  0.00001$), was also numerically analyzed. This is shown in Fig.~\ref{Fin}, where  1000 realizations with $\gamma = 0.5$ were  employed. This figure shows oscillations for the nonlinear inertial model which  are due to the fact that  for small noises, the deterministic part dominates. Those oscillations were seen to decay rapidly when  noise was large. Notice from the plot, that  for longer times the MSD profile, in both, overdamped and underdamped models, is not a constant but rather it linearly grows in time. It seems to be constant since $D_0$ is very small. In the same figure, Eq. (\ref{eccMSDmode}) (valid in the overdamped limit) is plotted in dashed lines. One can see its agreement with simulations for all times.

\section{Discussion}
\label{secdiss}
Some observations are in order. First. Through a rigorous  analysis one finds that the center of mass of a non-reciprocal system is no longer a dynamically decoupled translational coordinate. Its right zero mode remains the rigid translation $\mathbf 1$, but its corresponding left zero mode becomes nonuniform. As a result, the system's internal relaxation contributes to the finite-time MSD, whereas its long-time slope is fixed by the normalized left zero mode. In the reciprocal limit, the two coordinates coincide and the usual Rouse result $D_0/N$ is recovered.

Second. Through a rigorous  analysis we proved that the same picture remains valid for the underdamped case. We observed that inertia changes the system's ballistic at intermediate-time regimes, but  not its asymptotic diffusivity.  Third. For finite $l_0$, the relation $\mathbf f_R(\mathbf b)=q\mathbf f_L(\mathbf b)$ makes the internal forces to cancel exactly in the weighted translational coordinate. Fourth. For the smooth regularized force dynamics, it was proved that the system's bond second moments remain bounded. Note that $l_0$ can still change the bond statistics and the relaxation time, and therefore it can strongly modify the transient MSD. This also explains why long pre-asymptotic regimes may appear even though the final behavior is diffusive, and finally, a connection with an active Rouse chain tangentially driven was established. The current long-time effective diffusion coefficient of  a no-reciprocal Rouse chain contains in the limit $N \to \infty$, the reported effective coefficient in \cite{Peterson}, thus proving that our result is valid for any  number of particles. 

\section{Conclusions}
\label{sec:conclusions}

In summary, we studied the stochastic dynamics of the center-of-mass of a free non-reciprocal Rouse chain  both, overdamped and underdamped. For either model, it was found that the system's MSD linearly grows in time. For the overdamped case, an exact transient expression was obtained. By doing so, a spectral analysis showed that its non-symmetric drift matrix has one translational zero mode and $N-1$ stable internal modes. Its normalized left zero mode allowed us to obtain the system's long time effective diffusion coefficient. This coefficient indicated that diffusion is enhanced as non-reciprocity grows. 

We also found that the same diffusion coefficient holds for the underdamped linear model and for the finite-$l_0$ (finite bond length) nonlinear one (which is actually less studied in the literature). In the latter case, it was observed that the directional bond forces cancel exactly, and that the proposed smooth regularized force has bounded second moments. Brownian dynamics simulations showed that the current theory and simulations coincide, thus validating our findings.

\section{Acknowledgments}
M S thank the Mexican working community whose taxes support this research through Secretaria de Ciencia, Humanidades, Tecnologia e Innovacion (Secihti).

\section{Data availability statement}
The data that support this study are available  at 
https://espinozagroup.weebly.com/data-availability.html 

\appendix

\section{Diagonal symmetrization and positivity of the spectrum}
\label{apendsym}

Here we show explicitly that the drift matrix is diagonally similar to a real symmetric matrix and that its spectrum contains one zero eigenvalue and $N-1$ positive eigenvalues.

\subsection{Quadratic form and positivity of the internal spectrum}

Define the symmetrized elastic matrix $\mathbf K_s:=\mathbf S^{-1}\mathbf K\mathbf S
 =\gamma\mathbf A_s$. Let $\mathbf{x}\in\mathbb{R}^N$. Since $\mathbf K_s$ is real and symmetric, its spectral properties are governed by the associated quadratic form:
\begin{align}
 \mathbf{x}^T \mathbf{K}_s \mathbf{x}
 &= k_R x_1^2 + \sum_{n=2}^{N-1}(k_L+k_R)x_n^2 + k_L x_N^2 \nonumber \\
 &\quad - 2\sqrt{k_L k_R}\sum_{n=1}^{N-1}x_n x_{n+1}.
\end{align}
By regrouping terms, this quadratic form can be expressed as a sum of squares:
\begin{equation}
 \mathbf{x}^T \mathbf{K}_s \mathbf{x} = \sum_{n=1}^{N-1}\left(\sqrt{k_R}\,x_n - \sqrt{k_L}\,x_{n+1}\right)^2 \ge 0.
\end{equation}
Thus, $\mathbf{K}_s$ and consequently the drift matrix $\mathbf{A}_s = \mathbf{K}_s/\gamma$  is positive semidefinite, hence establishing that all
the eigenvalues satisfy $\lambda_n \ge 0$. 

Equality $\mathbf{x}^T \mathbf{K}_s \mathbf{x} = 0$ holds if and only if each term in the sum vanishes simultaneously, namely
\begin{equation}
 \sqrt{k_R}\,x_n - \sqrt{k_L}\,x_{n+1} = 0, \qquad n=1,\dots,N-1,
\end{equation}
which yields the recurrence relation
\begin{equation}
 x_{n+1} = \sqrt{\frac{k_R}{k_L}}\,x_n = \sqrt{q}\,x_n.
\end{equation}
This condition restricts the kernel $\ker(\mathbf{A}_s)$ to a unique one-dimensional subspace spanned by the zero-mode eigenvector. 

Because the kernel is one-dimensional, the Rayleigh--Ritz characterization on its orthogonal complement gives the spectral gap
\begin{equation}
 \lambda_2
 =
 \min_{\substack{\mathbf{x}\perp\ker(\mathbf A_s)\\ \|\mathbf x\|=1}}
 \mathbf{x}^T\mathbf A_s\mathbf{x}
 >0.
\end{equation}
All remaining eigenvalues satisfy $\lambda_n\ge\lambda_2>0$ for $n=2,\ldots,N$. Therefore, $\mathbf A_s$ possesses exactly one zero eigenvalue and $N-1$ strictly positive eigenvalues. Because $\mathbf A$ and $\mathbf A_s$ are related by a diagonal similarity transformation, their spectra coincide, proving the strict positivity of the internal relaxation spectrum of $\mathbf A$.

\section{Details of the second-moment derivation}
\label{apelyap}

Let us derive the explicit solution of the differential Lyapunov equation
\begin{equation}
 \dot{\mathbf{C}}(t)=-\mathbf{A}\mathbf{C}(t)-\mathbf{C}(t)\mathbf{A}^T+2D_0\mathbf{I}.
\end{equation}
Define
\begin{equation}
 \widetilde{\mathbf{C}}(t):=e^{\mathbf{A}t}\mathbf{C}(t)e^{\mathbf{A}^T t}.
\end{equation}
Differentiating,
\begin{align}
 \frac{d}{dt}\widetilde{\mathbf{C}}(t)
 &=
 \mathbf A e^{\mathbf A t}\mathbf{C}(t)e^{\mathbf A^Tt}
 +e^{\mathbf A t}\dot{\mathbf{C}}(t)e^{\mathbf A^Tt}
 \nonumber\\
 &\quad
 +e^{\mathbf A t}\mathbf{C}(t)\mathbf A^T e^{\mathbf A^Tt}
 \\
 &=
 e^{\mathbf A t}
 \left[
 \mathbf A\mathbf{C}(t)
 +\dot{\mathbf{C}}(t)
 +\mathbf{C}(t)\mathbf A^T
 \right]
 e^{\mathbf A^Tt}
 \\
 &=
 2D_0e^{\mathbf A t}e^{\mathbf A^Tt}.
\end{align}
Integrating from $0$ to $t$ gives
\begin{equation}
 \widetilde{\mathbf{C}}(t)-\widetilde{\mathbf{C}}(0)
 =
 2D_0\int_0^t e^{\mathbf{A}\tau}e^{\mathbf{A}^T \tau}\,d\tau.
\end{equation}
Since $\widetilde{\mathbf{C}}(0)=\mathbf{C}_0$, we obtain
\begin{equation}
 \widetilde{\mathbf{C}}(t)=\mathbf{C}_0+2D_0\int_0^t e^{\mathbf{A}\tau}e^{\mathbf{A}^T \tau}\,d\tau.
\end{equation}
Multiplying from the left by $e^{-\mathbf{A}t}$ and from the right by $e^{-\mathbf{A}^T t}$ gives
\begin{equation}
 \mathbf{C}(t)=e^{-\mathbf{A}t}\mathbf{C}_0e^{-\mathbf{A}^T t}
 +2D_0\int_0^t e^{-\mathbf{A}(t-\tau)}e^{-\mathbf{A}^T(t-\tau)}\,d\tau.
\end{equation}
A change of variables $u=t-\tau$ yields
\begin{equation}
 \mathbf{C}(t)=e^{-\mathbf{A}t}\mathbf{C}_0e^{-\mathbf{A}^T t}
 +2D_0\int_0^t e^{-\mathbf{A}u}e^{-\mathbf{A}^T u}\,du,
\end{equation}
which is the form used in the main text.

\section{Left zero mode and closed form of the effective diffusivity}
\label{appleftzero}
\subsection{Recurrence relation for the left zero mode}
The left zero mode satisfies $\mathbf{A}^T\boldsymbol{\psi}_1=\mathbf{0}$. For $N=2$, the boundary equation immediately gives $\psi_{1,2}=q\psi_{1,1}$, so the result below follows without an interior recurrence. Assume henceforth $N\ge3$. Writing $\boldsymbol{\psi}_1=(\psi_{1,1},\dots,\psi_{1,N})^T$, the interior recurrence for $2\le n\le N-1$ is
\begin{equation}
 -k_L\psi_{1,n+1}
 +(k_L+k_R)\psi_{1,n}
 -k_R\psi_{1,n-1}
 =0.
\end{equation}
We seek a geometric solution $\psi_{1,n}\propto \rho^{n-1}$. Substituting it into the previous equation gives the characteristic equation
\begin{equation}
 k_L\rho^2-(k_L+k_R)\rho+k_R=0,
\end{equation}
which factors as
\begin{equation}
 (\rho-1)(k_L\rho-k_R)=0.
\end{equation}
Thus the roots are $\rho=1$ and $\rho=q=k_R/k_L$, so the interior solution is a linear combination of a constant sequence and $q^{n-1}$. The left boundary equation $k_R\psi_{1,1}-k_L\psi_{1,2}=0$, requires $\psi_{1,2}=q\psi_{1,1}$. For $q\neq1$, this eliminates the constant component and leaves $\psi_{1,n}\propto q^{n-1}$. The right boundary equation is then automatically satisfied. For $q=1$, the left zero mode is uniform.

\subsection{Normalization}
By imposing the biorthogonal normalization $\boldsymbol{\psi}_1^T\boldsymbol{\phi}_1=1$ and $\boldsymbol{\phi}_1:=(1,\dots,1)^T$, we obtain $\sum_{n=1}^N \psi_{1,n}=1$. Therefore,
\begin{equation}
 \psi_{1,n}=\frac{q^{n-1}}{\sum_{m=0}^{N-1} q^m},
 \qquad n=1,\dots,N.
\end{equation}
For $q\neq 1$,
\begin{equation}
 \sum_{m=0}^{N-1} q^m=\frac{1-q^N}{1-q}.
\end{equation}
Hence
\begin{equation}
 \psi_{1,n}=\frac{1-q}{1-q^N}\,q^{n-1},
 \qquad q\neq 1.
\end{equation}

\subsection{Calculating the norm of the left zero mode}
We  directly obtain that
\begin{equation}
 \boldsymbol{\psi}_1^T\boldsymbol{\psi}_1
 =
 \sum_{n=1}^N \psi_{1,n}^2
 =
 \left(\frac{1-q}{1-q^N}\right)^2 \sum_{n=1}^N q^{2(n-1)}.
\end{equation}
For $q\neq 1$,
\begin{equation}
 \sum_{n=1}^N q^{2(n-1)} = \frac{1-q^{2N}}{1-q^2}.
\end{equation}
Therefore,
\begin{equation}
 \boldsymbol{\psi}_1^T\boldsymbol{\psi}_1
 =
 \frac{(1-q)^2(1-q^{2N})}{(1-q^N)^2(1-q^2)}.
\end{equation}
Furthermore, by using
\begin{equation}
 1-q^{2N}=(1-q^N)(1+q^N),
 \qquad
 1-q^2=(1-q)(1+q),
\end{equation}
we arrive at
\begin{equation}
 \boldsymbol{\psi}_1^T\boldsymbol{\psi}_1
 =
 \frac{(1-q)(1+q^N)}{(1+q)(1-q^N)}.
\end{equation}
Finally,
\begin{equation}
 D_\eff(q,N)=D_0\,\boldsymbol{\psi}_1^T\boldsymbol{\psi}_1
 = D_0\,\frac{(1-q)(1+q^N)}{(1+q)(1-q^N)}.
\end{equation}
In the reciprocal limit $q\to 1$, the normalized left zero mode becomes uniform and one recovers $D_\eff\to D_0 /N$.

\section{Initial conditions and their role in the MSD}
The initial state of the system enters the exact MSD only through the uncentered second-moment matrix $\mathbf C_0$ defined in Eq.~\eqref{ecuaC0}. Note that, for a zero-mean state, $\mathbf C_0$ coincides with the covariance. For a deterministic initial condition $\mathbf r(0)=\mathbf x_0$, one has $\mathbf{C}_0=\mathbf x_0\mathbf x_0^T$, and the corresponding contribution to Eq.~\eqref{eccMSDexacto} reduces to
\begin{equation}
 \mathbf{c}^T(e^{-\mathbf A t}-\mathbf I)\mathbf x_0\mathbf x_0^T
 (e^{-\mathbf A^T t}-\mathbf I)\mathbf{c}
 =
 \left[\mathbf{c}^T(e^{-\mathbf A t}-\mathbf I)\mathbf x_0\right]^2.
\end{equation}
A zero-mean Gaussian state with covariance $\sigma_0^2\mathbf I$ instead gives $\mathbf{C}_0=\sigma_0^2\mathbf I$, while sampling the stable component from the invariant Gaussian measure in Eq.~\eqref{equaststatsecmom}, makes the internal statistics stationary from the initial time.

Thus, the choice of $\mathbf{C}_0$ modifies the short- and intermediate-time MSD but not its asymptotic slope, provided the initial second moment is finite. The long-time growth remains controlled by the translational zero mode and is characterized by $D_{\eff}$.

\section{Confinement of the regularized nonlinear chain}
\label{appconfinement}
We now show that the smooth regularization in Eq.~\eqref{eqregularizedforce} gives uniformly bounded bond second moments. To do so, define the positive weights $w_n:=q^{n-1}$ and let $\mathbf M:=\mathrm{diag}(w_1,\ldots,w_N)\otimes\mathbf I_d$. Since $w_{n+1}=qw_n$, the regularized internal force satisfies
\begin{equation}
 \mathbf M\mathbf F(\mathbf r)=-\nabla_{\mathbf r}V_\varepsilon(\mathbf r),
\end{equation}
with
\begin{align}
 V_\varepsilon(\mathbf r)
 &:={\sum}_{j=1}^{N-1} w_jk_R\,\Phi_\varepsilon(\mathbf b_j),\\
 \Phi_\varepsilon(\mathbf b)
 &:=\frac12\|\mathbf b\|^2-l_0\sqrt{\|\mathbf b\|^2+\varepsilon^2}.
\end{align}
For large bond extensions this potential grows quadratically. In particular, there exist constants $c_0,C_0>0$ such that
\begin{equation}
 \Phi_\varepsilon(\mathbf b)\ge c_0\|\mathbf b\|^2-C_0,
 \qquad
 \mathbf b\cdot\nabla\Phi_\varepsilon(\mathbf b)
 \ge c_0\|\mathbf b\|^2-C_0.
 \label{eqcoercivePhi}
\end{equation}

Let $\mathbf X$ and $\mathbf U$ be the weighted translational variables defined in Sec.~\ref{secunderdampedloti}. We introduce the internal coordinates $\mathbf x_n:=\mathbf r_n-\mathbf X$ and $\mathbf u_n:=\mathbf v_n-\mathbf U$. They satisfy $\sum_n w_n\mathbf x_n=\sum_n w_n\mathbf u_n=\mathbf0$, while their bond vectors are still $\mathbf b_j=\mathbf x_{j+1}-\mathbf x_j$. On this weighted zero-mean subspace, a discrete Poincar\'e inequality gives
\begin{equation}
 \sum_{n=1}^{N}w_n\|\mathbf x_n\|^2
 \le C_P\sum_{j=1}^{N-1}\|\mathbf b_j\|^2
\end{equation}
for some finite $C_P$ depending only on $N$ and $q$.

Consider now the internal energy
\begin{equation}
 H:=\frac m2\sum_{n=1}^{N}w_n\|\mathbf u_n\|^2+V_\varepsilon+C_H,
\end{equation}
where $C_H$ is chosen so that $H\ge1$, and the cross term
\begin{equation}
 G:=\sum_{n=1}^{N}w_n\,\mathbf x_n\cdot\mathbf u_n.
\end{equation}
For the generator $\mathcal L$ of the internal Langevin dynamics, cancellation between the elastic work and $dV_\varepsilon$ gives
\begin{equation}
 \mathcal LH
 =-\gamma\sum_{n=1}^{N}w_n\|\mathbf u_n\|^2+C_T,
\end{equation}
where $C_T<\infty$ is the constant noise-injection term. Moreover,
\begin{equation}
 \mathcal LG
 =\sum_{n=1}^{N}w_n\|\mathbf u_n\|^2
 -\frac1m\,\mathbf x\cdot\nabla V_\varepsilon
 -\frac\gamma m G.
\end{equation}
Using Eq.~\eqref{eqcoercivePhi}, the discrete Poincar\'e inequality, and Young's inequality, $\theta>0$ can be chosen sufficiently small so that the Lyapunov function $\mathcal V:=H+\theta G$ is equivalent to
$1+\sum_n\|\mathbf u_n\|^2+\sum_j\|\mathbf b_j\|^2$ and satisfies
\begin{equation}
 \mathcal L\mathcal V\le-c\mathcal V+C
 \label{eqFosterLyapunov}
\end{equation}
for constants $c,C>0$. Dynkin's formula and Gronwall's inequality then yield
\begin{equation}
 \sup_{t\ge0}
\left\langle \mathcal V(t)\right\rangle
<\infty,
\end{equation}
and therefore
\begin{equation}
 \sup_{t\ge0}\left\langle\|\mathbf b_j(t)\|^2\right\rangle<\infty,
 \qquad j=1,\ldots,N-1.
\end{equation}
Equation~\eqref{eccbond_confinement} therefore follows for the smooth regularized dynamics. For the singular law $\varepsilon=0$, the same coercive estimate holds on the collision-free domain. However, it does not prove that a trajectory cannot reach $\mathbf b_j=\mathbf0$, so the unregularized statement remains conditional on global collision-free existence.

\bibliography{biblio}

\end{document}